\documentclass[prl,twocolumn,groupedaddress,superscriptaddress]{revtex4-1}

\usepackage{eurosym}
\usepackage{amsfonts}
\usepackage{amsmath}
\usepackage{amssymb}
\usepackage{graphicx}
\usepackage{float}

\begin{document}
	
\newcommand{\TN}{$T_N$}
\newcommand{\Tc}{$T_c$}
\newcommand{\msr}{$\mu$SR}

\title{Relevance of magnetism to cuprate superconductivity: Lanthanides versus charge-compensated cuprates}

\author{Amit Keren}
\affiliation{Department of Physics, Technion - Israel Institute of Technology, Haifa, 3200003, Israel}

\author{Wayne Crump}
\affiliation{Robinson Research Institute, and MacDiarmid Institute, P.O. Box 33436 Lower Hutt, New Zealand}

\author{Ben P. P. Mallett}
\affiliation{The Photon Factory, Department of Physics, The University of Auckland, 38 Princes St, Auckland 1010, New Zealand}

\author{Shen V. Chong}
\affiliation{Robinson Research Institute, and MacDiarmid Institute, P.O. Box 33436 Lower Hutt, New Zealand}

\author{Itai Keren}
\affiliation{Paul Scherrer Institute, CH 5232 Villigen PSI, Switzerland}

\author{Hubertus Luetkens}
\affiliation{Paul Scherrer Institute, CH 5232 Villigen PSI, Switzerland}

\author{Jeffery L. Tallon}
\affiliation{Robinson Research Institute, and MacDiarmid Institute, P.O. Box 33436 Lower Hutt, New Zealand}

\date{\today }

\begin{abstract}

We address what seemed to be a contradiction between the lanthanide series REBa$_2$Cu$_3$O$_y$ (RE123) and the charge-compensated series (Ca$_{x}$La$_{1-x}$)(Ba$_{1.75-x}$La$_{0.25+x} $)Cu$_{3}$O$_{y}$ (CLBLCO) regarding the superexchange ($J$) dependence of the maximum superconductivity (SC) critical temperature $T_c^{max}(J)$; RE and $x$ are implicit variables. This is done by measuring the N\'{e}el temperature and the temperature dependence of the magnetic order parameter for RE=Nd, Sm, Eu, Gd, Dy, Yb, Y, and for Y(BaSr)Cu$_3$O$_y$, at various very light dopings. The doping is determined by thermopower, and the magnetic properties by muon spin rotation. We find that the normalized-temperature dependence of the order parameter is identical for all RE123 in the undoped limit (with the exception of Gd123) implying identical out-of-plane magnetic coupling. The extrapolation of $T_N$ to zero doping suggests that, despite the variations in ionic radii, $J$ varies too weakly in this system to test the relation between SC and magnetism. This stands in contrast to CLBLCO where both $T_c^{max}$ and $T_N^{max}$ vary considerably in the undoped limit, and a positive correlation between the two quantities was observed.

\end{abstract}

\maketitle

\footnotetext[1]{Correspondence should be addressed to A.K. (email: keren@physics.technion.ac.il)}

\section{Introduction}

Recently, the group of Tallon \cite{MallettPRL13} measured the in plane super-exchange parameter $J$ in a series of samples similar to YBa$_2$Cu$_3$O$_y$, where Y was replaced by one of the lanthanides: La, Nd, Eu, Gd, Dy, Yb, Lu, or the Ba$_2$ was modified to BaSr. The measurements were done using two-magnon Raman scattering. The samples were prepared with as low doping ($p$) as possible, although the actual value was not determined. They found that as one progresses in the lanthanide series and the atomic number increases, $J$ also increases. They justify this $J$ increase by the famous lanthanide contraction where the atomic radius becomes smaller as the atomic number increases.
They also found anti-correlation between the maximum $T_c$ ($T_c^{max}$) of each family of materials, and $J$. The internal pressure (induced by substitution of isovalent ions of smaller size) seems to increase $J$ but decrease $T_c^{max}$.

The RE123 result stands in strong contrast to experiments on the charge-compensated compound (Ca$_{x}$La$_{1-x}$)(Ba$_{1.75-x}$La$_{0.25+x} $)Cu$_{3}$O$_{y}$ (CLBLCO) performed by the Keren group. The name ``charge-compensated" comes from the fact that Ca and Ba have the same valance and their replacement does not formally dope the system. However increasing $x$ shrinks the Cu-O-Cu distance and straightens the buckling angle \cite{OferPRB08}; the total amount of La in the chemical formula is constant. In CLBLCO $J$ and $T_c$ were measured for various values of $x$ and $y$. It was found that $J$ in the parent and doped compounds and $T_c^{max}$ are correlated; the stronger the magnetic interactions the higher $T_c^{max}$ is. The measurements were done with muon spin rotation ($\mu$SR) \cite{OferPRB06}, Raman scattering \cite{WulferdingPRB14}, angle resolved photoemission \cite{DrachuckPRB14}, and resonant inelastic x-ray scattering \cite{EllisPRB15} and all methods agree qualitatively.  The RE123 results are also in contradiction with external pressure experiments on Y123 as pointed out by Tallon and co-workers \cite{MallettPRL13}. External pressure raises $T_c^{max}$ and $J$ simultaneously.

\begin{figure*}[h!t]
	\includegraphics[trim=0cm 0cm 1cm 0cm, clip=true,width=18cm]{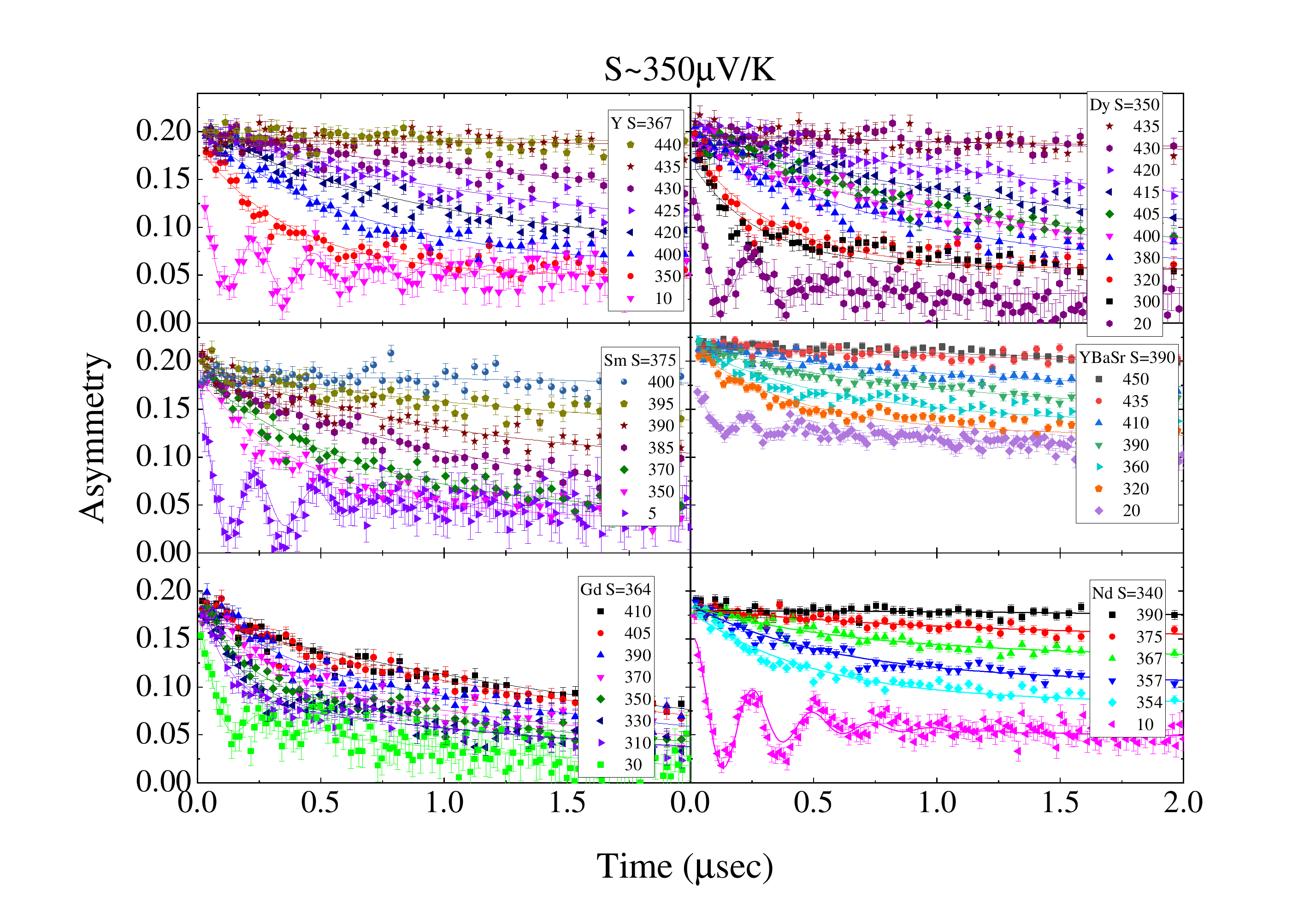}
	\caption{\textbf{Raw data of lightly doped samples.} The \msr\ asymmetry as a function of time for different RE123 samples with Seebeck coefficient $S\sim 350$~$\mu$V/K. The sample name, exact $S$, and temperatures are written in the labels. Results from temperatures near \TN\ and deep in the ordered state are presented. }
	\label{RawData350}
\end{figure*}

An attempt was made to resolve the contradiction using new two-magnon Raman scattering measurements \cite{Mullner18}. In this experiment only samples that are prepared under the same conditions, and with the doping determined by thermopower, where remeasured. It was found that within experimental uncertainty the RE=Y, Dy, Gd, and Sm have the same two-magnon Raman peak  frequency. The RE=Nd has a peak at substantially lower energy than its counterparts. This indicates that at least among the first four superconducting families $J$ is not changing appreciably with lanthanide substitution.

In this manuscript we address the same discrepancy from the perspective of magnetic measurements. We apply the $\mu$SR technique to 27 samples with different RE compositions and doping, including the Y(BaSr)Cu$_3$O$_y$. The doping is determined from the thermopower Seebeck coefficient ($S$) \cite{ObertelliPRB92}. For each sample we measure the N\'{e}el temperature ($T_N$) and the muon spin angular rotation frequency $\omega$ as a function of temperature. Since $T_N$ is set by both in-plane and out-of-plane coupling $J$ and $J_{\perp}$ respectively, two measured quantities are required to determine both couplings. These quantities are $T_N$ and the order parameter $\sigma(T)=\omega(T) / \omega_0$, where $\omega_0$ is the muon spin rotation frequency at $T \rightarrow 0$ \cite{OferPRB06}. This type of analysis works best in the fully undoped case which is described by the 3D Heisenberg Hamiltonian. But, since it is not clear if the samples are completely undoped, we perform measurements as a function of doping and extrapolate to zero doping.

\begin{figure*}[h!t]
	\includegraphics[trim=0cm 0cm 1cm 0cm, clip=true,width=18cm]{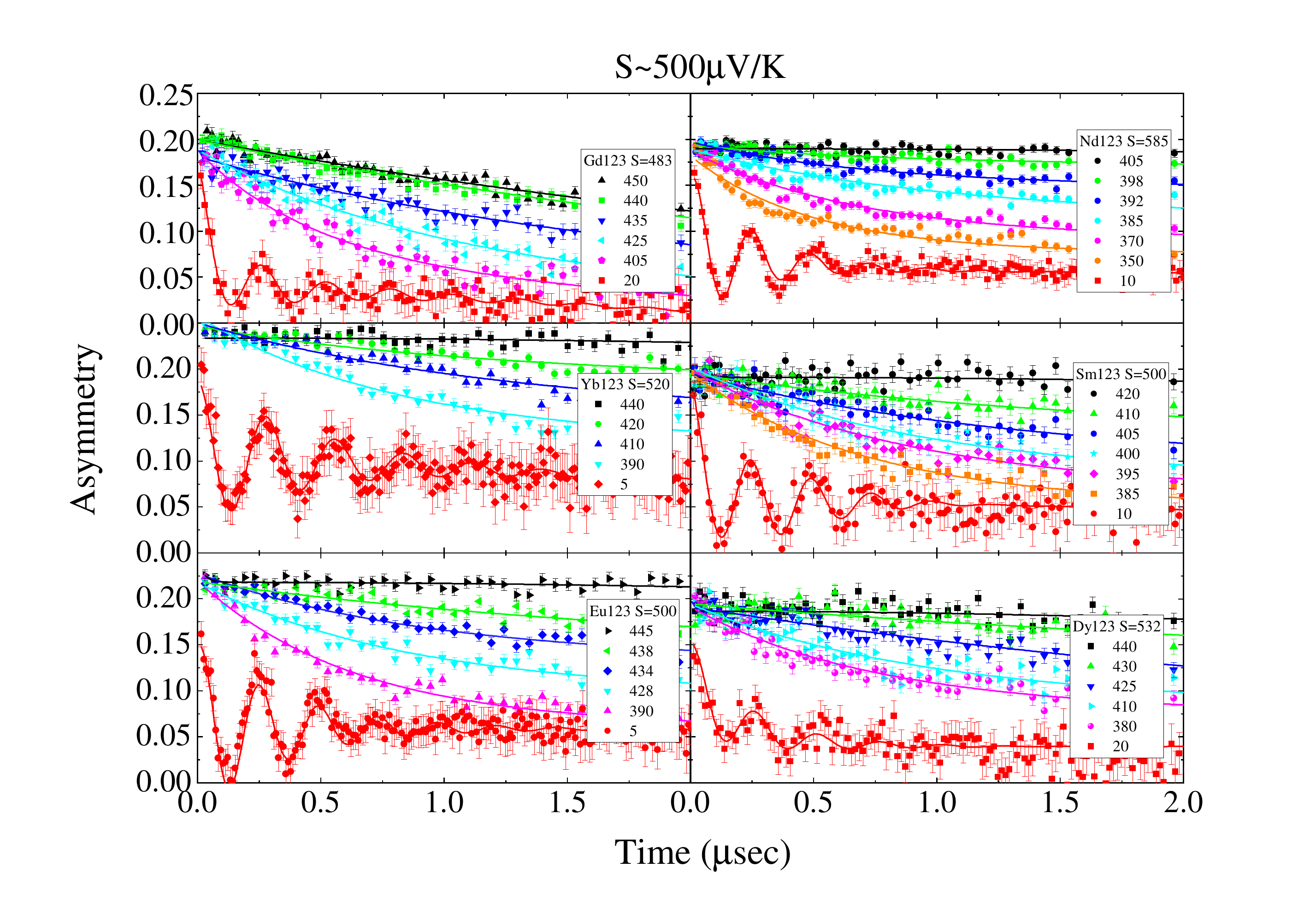}
	\caption{\textbf{Raw data of nearly undoped samples.} The \msr\ asymmetry as a function of time for different RE123 samples with Seebeck coefficient $S\sim 500$~$\mu$V/K. The sample name, exact $S$, and temperatures are written in the labels. Results from temperatures near \TN\ and deep in the ordered state are presented. }
	\label{RawData500}
\end{figure*}

\section{Experimental Aspects}

The RE123 and  Y(BaSr)Cu$_3$O$_y$ samples are prepared by solid-state reaction at ambient pressure \cite{GilioliIJMB00}. Each sample is an agglomerate of single crystals of sizes up to $100$~$\mu$m pressed into a pellet typically $10$~mm in diameter and $1$~mm thick. The doping of the crystals is set by argon annealing at $T \sim 650$~C, followed by quenching into liquid nitrogen. The doping is determined from the room-temperature thermopower Seebeck coefficient, S(290), measured on the same samples used for the \msr\ measurements prior to the beam time. The \msr\ experiments are done on the GPS beam line at Paul Scherrer Institute using a closed cycle refrigerator which provides a temperature range of $5$ to $500$~K. Fine temperature scans where done close to the magnetic phase transition and between $5$ and $200$~K; above $200$~K muon diffusion sets in and hinders detailed data analysis but still allows the determination of \TN\ \cite{KerenPRB93}. The samples are cooled in zero field and the muon polarization as a function of time is determined via the asymmetry in decay positrons.

Raw data from the various RE123 systems with $S \sim 350$~$\mu$V/K are presented in Fig.~\ref{RawData350}. The exact value of $S$ is written in each panel. For all samples, apart from Gd123, there is a temperature high enough that the asymmetry does not relax on the time scale presented in the figure.  In all cases the asymmetry develops a strong relaxation within a temperature range of $10$~K below \TN. In the Gd case the asymmetry increases its relaxation from a high temperature saturated value. A finite, temperature independent relaxation rate at high temperatures, in samples containing Gd is ubiquitous (e.g., Ref.~\cite{ShekharPNAS18}). In all cases, at very low temperatures oscillations develop indicating an ordered state of the material with a site-average magnetic field at the muon site larger than its fluctuation from site to site. The oscillation frequency is similar in all samples. The signal from YBaSr indicates that only part of this sample is actually magnetic.

\begin{figure*}[h!t] 
	\includegraphics[trim=0cm 0cm 0cm 0cm, clip=true,width=17cm]{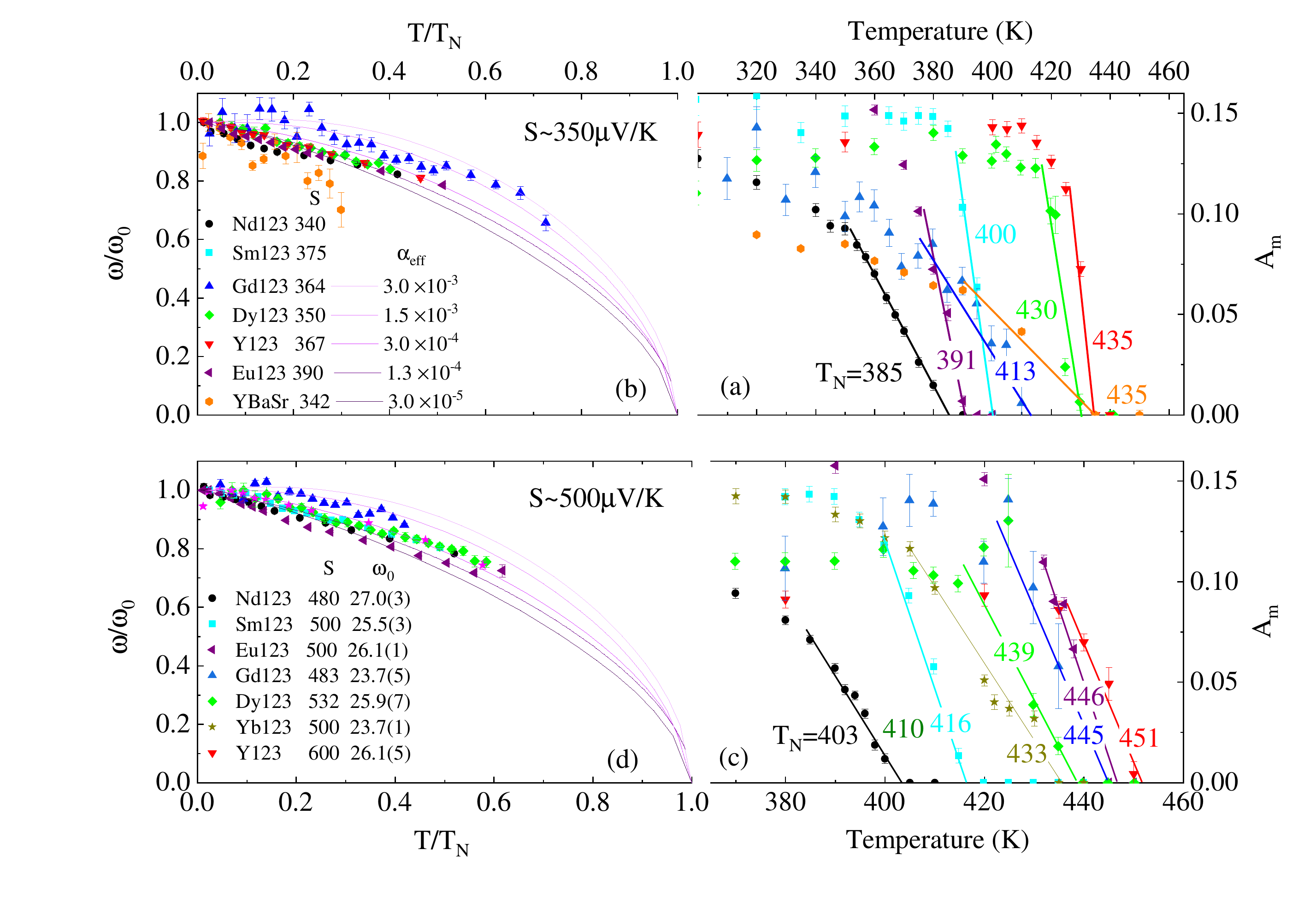}
	\caption{\textbf{Fit results.} For samples with $S\sim 350$~$\mu$V/K: (a) the magnetic asymmetry [see Eq.~\ref{FitEquation}] as a function of temperature. The solid lines are linear fits near the phase transition defining \TN, which is written next to each line. (b) Symbols are the order parameter evaluated by the normalized muon rotation frequency $\omega/\omega_0$, where $\omega_0=\omega(T \rightarrow 0)$, as a function of temperatures. The solid lines represent a calculation of an anisotropic 3D-Heisenberg model for different effective anisotropy parameter $\alpha_{eff}$ as described in the text. For the $S\sim 500$~$\mu$V/K samples: (c) and (d) are the same as (a) and (b) respectively. The values of $\omega_0$ in MRad/sec for the high $S$ samples are given in (d).}
	\label{Analysis}
\end{figure*}
 
Results from a representative set of RE123 with $S \sim 500$~$\mu$V/K are shown in Fig.~\ref{RawData500}.  Again, the relaxation in the Gd case is not zero at high temperatures but it saturates. For all families, the relaxation increases over a narrow temperature range (relative to the $T_N$). At low temperatures asymmetry oscillations develop. 

The time-dependent $\mu$SR asymmetry $A(t)$ data is analyzed with the function 
\begin{multline}
A(t)=A_{n}\exp\left( -(\Delta t)^\alpha \right)+ \\ A_{m}\left[ \exp\left( -t/T_{\parallel} \right) +R \exp\left( -(t/T_{\perp})^\gamma \right) \cos(\omega t)  \right].
\label{FitEquation}
\end{multline}
In this function $A_n$ represents the non-magnetic fraction of the sample, and $\Delta$ the relaxation rate of the muon spin in this part of the sample. $A_m$ is proportional to the magnetic fraction of the samples. $\alpha=2$ except for the Gd samples where $\alpha=1$ above room temperature. $\gamma = 1$ apart from Gd and Sm $S=500$ samples where the values $\gamma = 0.5$ and $2$ respectively provide the best fit. $T_{\parallel}$ and $T_{\perp}$ are the muon spin relaxation times in the direction of the local field at the muon site and perpendicular to it, respectively. The relaxation rate $\Delta$ varies between samples but is kept fixed in the fit for each sample. In principle $R$ should be $2$ since there are two field components perpendicular to the muon spin compared to only one longitudinal component. In practice $R$ is a fit parameter. Also the total asymmetry should be shared at all temperatures. In practice it is shared for temperatures between $5$ and $200$~K and between $200$ to $470$~K separately. Finally, $\omega$ is the muon rotation frequency. It is set to zero when no oscillations are observed in the data, in which case $R$ is also set to zero and $T_\parallel$ has no directional association.

The relevant fit parameters are depicted in Fig.~\ref{Analysis}. Panel (a) shows the magnetic fraction $A_m$ as a function of temperature for samples with $S \sim 350$~$\mu$V/K. A straight line is fitted to the sharp rise in $A_m$ and the point of abscissa crossing defines \TN. The value of \TN\ varies between $385$ and $435$~K and is indicated next to each line. The sharpness of the phase transition also varies between families. The symbols in panel (b) show the temperature-dependent order parameter. Eu, Sm, Dy, and Y families have the same rate of order parameter reduction with increasing temperature. Gd has a smaller, and Nd and YBaSr have higher reduction rates than the common one. $\sigma(T)$ is a measure of the magnetic coupling anisotropy. The smaller $ \frac{d\sigma}{dT} $ at $T \rightarrow 0$ the more isotropic 3D-like is the magnetic system \cite{ArovasPRB886}.

The solid lines in panel (b) are the self-consistent Schwinger-boson mean-field theory calculations \cite{ArovasPRB886} of $\sigma(\alpha_{eff},t)$ where $t=T/J$, $\alpha_{eff}=z_{xy}\alpha_{xy}+z_{\perp}\alpha_{\perp} $, the $z$'s are the number of neighbors, $\alpha_{xy}$ is the in-plane anisotropy, and $\alpha_{\perp}=J_\perp / J $. Since RE123 has two types of $J_{\perp}$, this parameter represents an average perpendicular coupling. More details are given in Ref.~\cite{OferPRB06}. However, this model is valid for the Heisenberg Hamiltonian, and the samples presented in Fig.~\ref{Analysis}(b) are slightly doped. The analysis becomes more accurate as $S$ increases further.

Figure~\ref{Analysis} panels (c) and (d) also present $A_m$ and $\omega/\omega_0$ but for samples with $S \sim 500$~$\mu$V/K. In this case the lowest value of \TN\ is $400$~K and therefore the spread in \TN\ between different families is smaller. In addition, apart from Gd, all $\sigma (T)$ at $T \rightarrow 0$ nearly overlap and $\alpha_{eff}$ is on the order of $10^{-5}$. This result suggests that as the doping decreases the different families converge to the same magnetic behavior.

\section{Discussion}

\begin{figure}[tbph]
	\includegraphics[trim=0cm 0cm 0cm 0cm, clip=true,width=\columnwidth]{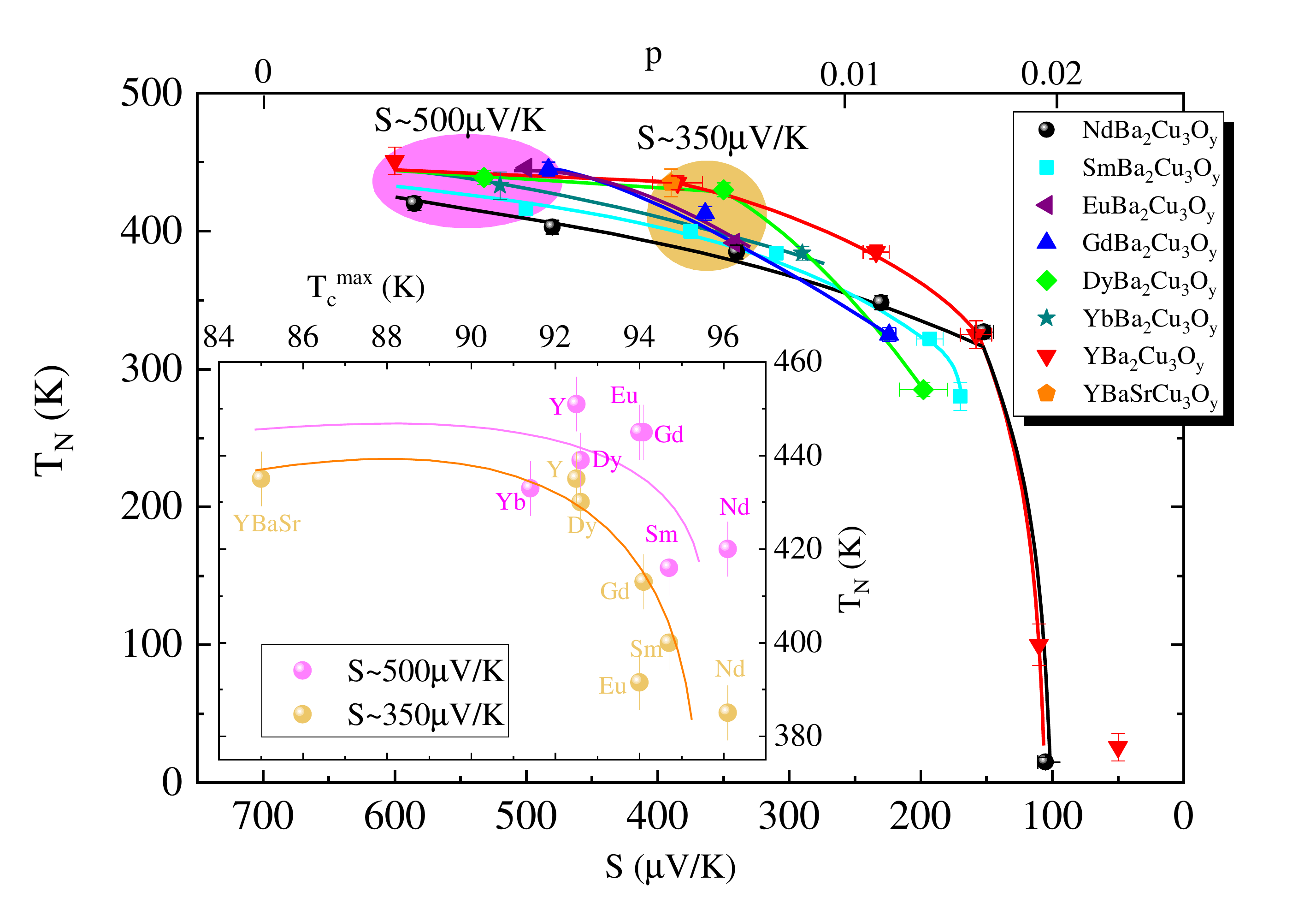}
	\caption{\textbf{N\'{e}el temperature summary.} \TN\ versus the thermoelectric power Seebeck coefficient $S$ as measured at room temperature for the different samples. The larger the $S$, the smaller the doping. A rough estimate of the doping $p$ is obtained from the relation $S=700\exp(-100p)$ (see text). As the doping decreases the variation in \TN becomes smaller. Two groups of samples are highlighted, each with similar value of $S$. The inset show $T_c^{max}$ of each family as a function of \TN. For $S\sim 350$~$\mu$V/K anti-correlation between $T_c^{max}$ and $T_N$ is observed for all samples excluding YBaSr. However, for $S\sim 500$~$\mu$V/K the variation in \TN\ is not systematic. }
	\label{Conclusions}
\end{figure}

Our main results are depicted in Fig.~\ref{Conclusions}. We present $T_N$ as a function of the thermopower $S$ in the lower abscissa for various RE123 families and Y(BaSr)Cu$_3$O$_y$; $S$ decreases with increasing doping and hence the reverse axis. For all families, $T_N$ increases with increasing $S$ (decreasing doping). For some of the families such as RE=Y, and Dy,  a saturation is clearly reached which reflects the fact that for these families doping is not changing at these high thermopower values. In RE=Gd and Eu it is not clear if saturation has been reached. For RE=Yb, Sm, and Nd it is clear that saturation has not been reached, and if it was possible to extract more oxygen from the sample, $T_N$ could have increased. In all families $T_N$ never exceeds $450$~K. This is particularly peculiar for the YBaSrCu$_3$O$_y$ where $J$ is larger by 10\%, according to new measurements \cite{Wulferding18}, and according to the original measurements \cite{MallettPRL13} in YBa$_{0.5}$Sr$_{1.5}$Cu$_3$O$_y$ by 50\%, than in RE=Y. Therefore, $T_N^{max}$ of YBaSrCu$_3$O$_y$ is expected to be higher than $495$~K, which is not the case. The solid lines in the figure are guides to the eye.  It is conceivable but not guaranteed that all these lines meet at $S \cong 700$, which would then correspond to zero doping. These lines suggest that $T_N^{max}$s for all examined families may be identical. Assuming that (i) all lines flatten by $S \cong 700$, (ii) that at an estimated doping level $p \cong 0.02$ $T_N$ drops to zero, and (iii) that the relation between $S$ and $p$ is exponential, then we may convert $S$ to $p$ using the relation $S=700\exp (-100p)$. Values of $p$ thus obtained are presented on the top abscissa.

A different way of looking at the same data is depicted in the inset of  Fig.~\ref{Conclusions}. Here we plot $T_N$ versus $T_c^{max}$ for each SC family at two, roughly fixed $S$, namely fixed doping. The room temperature thermopower has been shown to be an excellent correlate of the doped hole concentration, $p$, in units of holes/Cu \cite{ObertelliPRB92}. Closer to optimal doping and beyond, it is a highly sensitive and a precise measure of doping, but at very low doping it becomes increasingly uncertain as $p \rightarrow 0$. For this reason we bin our doping states separately into $S \sim 350$~$\mu$V/K and $S \sim 500$~$\mu$V/K. The corresponding data in Fig.~\ref{Conclusions} is highlighted. In both cases $p<0.01$ holes/Cu and we are confident that the doping state of the latter is less than that of the former. Beyond such broad categories it is impossible to read much into any variations within the $350$~$\mu$V/K or $500$~$\mu$V/K groups; both are extremely close to the undoped insulator and the variations in \TN\ temperature seen between the two groups show just how sensitive \TN\ is to small increments in doping near $p = 0$. Looking at the $S \sim 350$~$\mu$V/K and ignoring Y(BaSr)Cu$_3$O$_y$ it seems that there is anti-correlation between $T_c^{max}$ and $T_N$. But, as $S$ increases towards $S \sim 500$~$\mu$V/K (doping decreases) this anti-correlation weakens, and, again, the effect of lanthanide substitution on $T_N$ disappears. 

\begin{figure}[tbph]
	\includegraphics[trim=0cm 0cm 0cm 0cm, clip=true,width=\columnwidth]{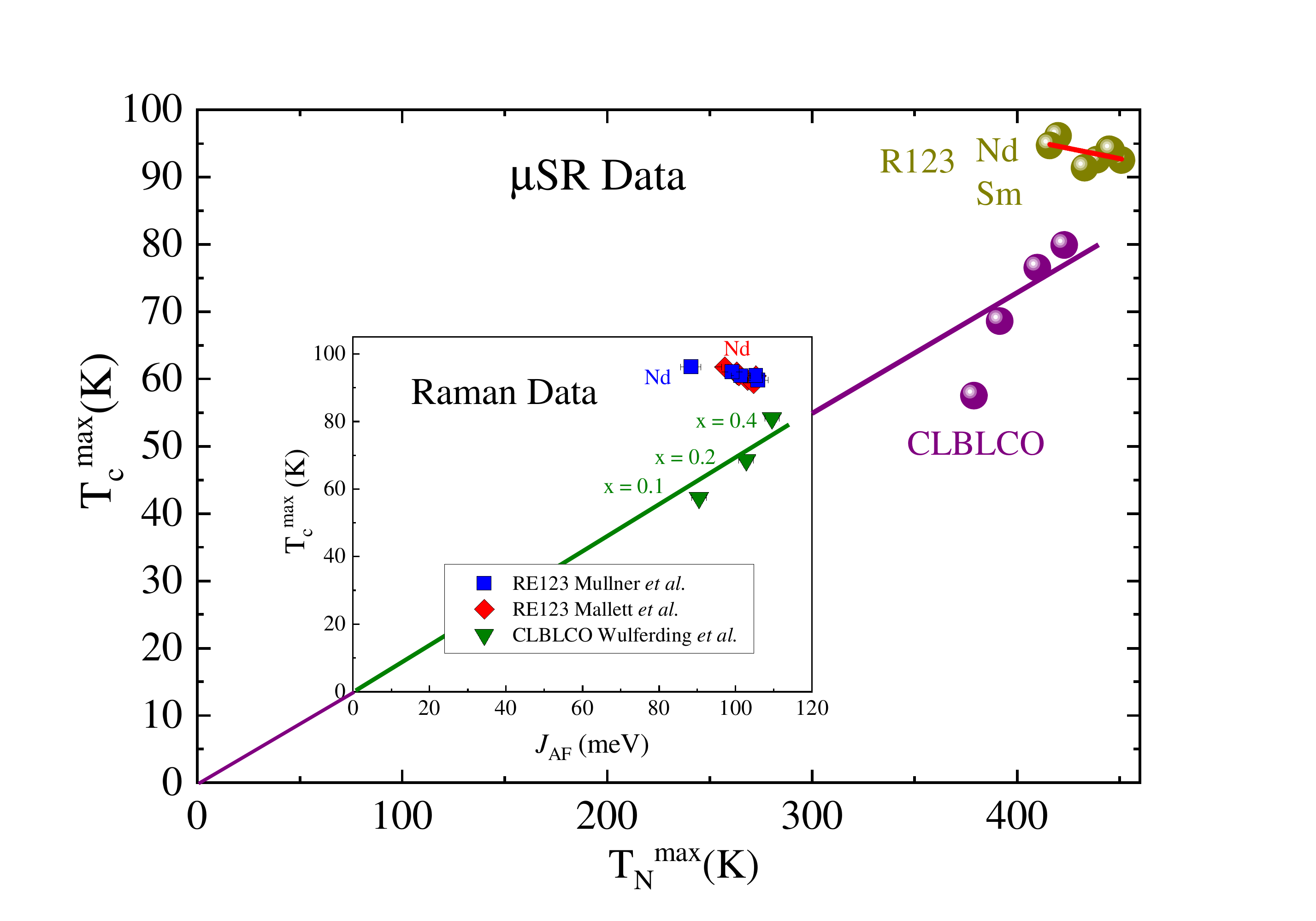}
	\caption{\textbf{RE123 versus CLBLCO.} The main panel depicts $T_c^{max}$ versus $T_N^{max}$ for the RE123 system as determined in the present experiment, and for CLBLCO from Ref.~\cite{OferPRB06}. $T_N^{max}$ is the maximum N\'{e}el temperature achieved by oxygen reduction (in contrast to the extrapolated value). The data is presented on a scale including the origin. The error bars are smaller than the symbols size. The red solid line has a slope of $-0.22\pm 0.16$ demonstrating that the dependence of $T_c^{max}$ on $T_N^{max}$ in RE123 is barely significant. The inset shows the Raman data from Mallett \textit{et al.} \cite{MallettPRL13}, Mullner \textit{et al.} \cite{Mullner18}, and Wulferding \textit{et al.} \cite{Wulferding18} on both systems again on a full scale. The Nd123 is emphasized due to lack of agreement between different groups. The solid lines through the origin serves to evaluate the agreement or disagreement of the data with proportionality between $T_c^{max}$ versus $T_N^{max}$.}
	\label{Summary}
\end{figure}

A comparison between RE123 and CLBLCO is presented in Fig.~\ref{Summary}. The main panel depicts the relations between $T_c^{max}$ and $T_N^{max}$, where $T_N^{max}$ is the maximum N\'{e}el temperature achieved for each family. For RE123 $T_c^{max}$ values are taken from Refs.~\cite{VealPhyscaC89,GuilaumeJPCM94,MallettPRL13}, and $T_N^{max}$ is the highest value of $T_N$ we managed to achieve for each family by oxygen reduction. For CLBLCO, a saturation of \TN\ is achieved for all $x$ values by underdoping, and there is no need to extrapolate \TN\ to zero doping. Both quantities are plotted on a full scale including the origin. In the RE123 case, a fit to a straight line gives a slope of $-0.22\pm 0.16$, namely, the error is similar to the value. This means that basically $T_c^{max}$ is independent of $T_N^{max}$. Furthermore, since Nd and Sm are slightly doped, in the ideal undoped case all points should be bunched together with no variation in $T_N^{max}$, as suggested from the extrapolation of the data in Fig.~\ref{Conclusions} to high $S$. In contrast, the CLBLCO points are well separated on both the $T_c^{max}$ and $T_N^{max}$ axis.

The inset shows data from the Raman measurements, also plotted on a full scale, but only for samples which are prepared under the same condition (ambient pressure) and measured by both M\"{u}llner \cite{Mullner18} and Mallett \cite{MallettPRL13}. Close examination of this data shows an anti-correlation between $T_c^{max}$ and $T_N^{max}$, however, there is disagreement on the Nd data point, and overall it seems that on a scale including the origin, neither $T_c^{max}$ nor $T_N^{max}$ changes enough in the R123 samples to allow for a proper examination of the relation between magnetism and superconductivity with this system. Again, in CLBLCO, both quantities change by more than 10\% and the experimental message is clearer. One possibility is that indeed $T_c^{max}$ scales with $J$, however a different interpretation is that CLBLCO is anomalous as discussed in Ref.~\cite{TallonPRB14}. There it was shown that, once the doping is determined using thermopower, the pseudogap shows a universal doping-dependent behavior, independent of composition, $x$. In contrast, $T_c^{max}$ appears to show an anomalous suppression which grows with decreasing $x$, thereby effectively reversing its correlation with $J$. The reasons for such a suppression are not apparent and, for example, NMR studies suggest it is not associated with disorder scattering \cite{AmitPRB10, CvitaincPRB14, AgrestiniJPCS14}. These are interesting model systems that deserve more study if systematic behavior is to be elucidated.

\section{Conclusions}

We characterize the magnetic properties of several different RE123 compounds, and Y(BaSr)Cu$_3$O$_y$ as a function of doping. In particular we focus on the N\'{e}el temperature and the reduction of the order parameter $\sigma=\omega/\omega_0$ as a function of temperature. It is possible (yet not essential) to extrapolate the data for each family to zero doping in a way where all the RE123 have the same magnetic properties. In particular they have the same \TN. This is quite surprising considering the changes in unit cell parameters \cite{GuilaumeJPCM94}. Similarly, within experimental errors all RE123 presented here have nearly identical $T_c^{max}$. Therefore, RE123 is not the system with which one would like to test the relation between superconductivity and magnetism. In contrast CLBLCO shows large variation in both quantities and indicates a positive correlation between magnetic properties and superconductivity.

\section{Acknowledgments}

The Technion team is supported by the Israeli Science Foundation (ISF) grant number 315/17 and by Technion RBNI Nevet program. The $\mu$SR work is based on experiments performed on the GPS instrument of the Swiss Muon Source S$\mu$S, Paul Scherrer Institute, Villigen, Switzerland.


%

\end{document}